\documentclass[amsfonts,aps,prl,twocolumn,showpacs,superscriptaddress]{revtex4-1}
\usepackage{amssymb}
\usepackage{amsmath}
\usepackage{graphicx}
\newcommand{\beq}{\begin{equation}}  
\newcommand{\beqa}{\begin{eqnarray}}  
\newcommand{\eeqa}{\end{eqnarray}}  
  
\newcommand{\ome}{\omega}

\usepackage{xcolor}

\begin{document}

\title{The emergence of classical behavior in magnetic adatoms }

\author{F. Delgado}
\affiliation{International Iberian Nanotechnology Laboratory (INL),
Av. Mestre Jos\'e Veiga, 4715-310 Braga, Portugal}

\author{S.Loth}
\affiliation{Max Planck Institute for the Structure and Dynamics of Matter, Hamburg, Germany}
\affiliation{Max Planck Institute for  Solid State Research, Stuttgart, Germany}

\author{M. Zielinski}
\affiliation{Institute of Physics, Faculty of Physics,
Astronomy and Informatics, Nicolaus Copernicus University,
Grudziadzka 5, 87-100 Torun, Poland}

\author{J. Fern\'andez-Rossier}
\affiliation{International Iberian Nanotechnology Laboratory (INL),
Av. Mestre Jos\'e Veiga, 4715-310 Braga, Portugal}
\altaffiliation[Permanent address: ]{Departamento de F\'{\i}sica Aplicada, Universidad de Alicante, San Vicente del Raspeig, 03690 Spain}
\email[Corresponding author:]{joaquin.fernandez-rossier@inl.int}

\date{\today}

\begin{abstract}

A wide class of  nanomagnets shows  striking  quantum behavior, known as quantum spin tunneling (QST): instead of two degenerate ground states with opposite magnetizations, a bonding-antibonding pair  forms, resulting in a splitting of the ground state doublet  with wave functions 
linear combination of two classically opposite magnetic states, leading to the quenching of  their magnetic moment.
%
%
  Here we study how QST is destroyed and classical behavior emerges in the case of 
  magnetic adatoms, as the strength of their coupling, either  to a bath or to each other,  is increased.
  Both  spin-bath and spin-spin coupling renormalize the QST splitting to zero
allowing the environmental decoherence to eliminate superpositions between classical states, leading to the emergence of spontaneous magnetization. 

\end{abstract}
\maketitle

Understanding how matter, governed by quantum mechanics at the atomic scale, behaves with classical rules at the macroscale is one of the fundamental open questions  in physics~\cite{Anderson_science_1972,Zurek_physics_today_91,Zurek_revmodphys_2003}.  
One of the most drastic manifestations of the quantum character is found when a system is prepared in a linear combination of two classically different states. In magnetic systems, such a quantum state results in the phenomenon of quantum spin tunneling (QST)~\cite{Gatteschi_Sessoli_book_2006}, inducing 
an energy splitting $\Delta_0$ between the two states with opposite magnetization and quenching its average magnetization.

Attending to the nature of their ground state,  nanoscale quantized spin systems  can be classified in two groups, see Figs. 1a,b. 
 Type C systems, such as single half-integer spins or antiferromagnetic
 chains of Ising coupled spins,  have two degenerate ground states with wave functions $|C_1\rangle$ and $|C_2\rangle$
that  correspond to states with opposite magnetizations.   Type Q  systems,  such as anisotropic single integer spins,   have a unique ground state $|\phi_G\rangle$,  as well as a first excited state $|\phi_X\rangle$, both satisfying
\begin{equation}
|\phi\rangle = |C_1\rangle+e^{i\theta}|C_2\rangle
\label{superp}
\end{equation}
where $\theta$ is a phase.  Whereas both correspond to quantum spins, only type Q  systems depart  radically from the classical picture of a nanomagnet because the quantum expectation  value $\langle \phi|\vec S|\phi\rangle$ of the atomic spin operator $\vec{S}$ vanishes
identically,  and not only in the statistical sense. 
  The two degenerate ground states of type C systems could be prepared in superposition states like  (\ref{superp}),  but  coupling to the environment  would rapidly lead to decoherence, restoring the  classical behavior with two ground states with opposite magnetization~\cite{Delgado_Rossier_prl_2012}.   In contrast, in type Q  systems the  coherent superposition is built-in dynamically,  and it is protected by the energy separation 
$\Delta_0$    between  $|\phi_G\rangle$ and $|\phi_X\rangle$, see Fig. 1b.
Examples of 
 type Q magnets are found in transition metal impurities in insulators~\cite{Abragam_Bleaney_book_1970}, spin color centers~\cite{Childress_Gurudev_science_2006}, magnetic molecules~\cite{Tsukahara_Noto_prl_2009} and single molecule magnets~\cite{Gatteschi_Sessoli_book_2006}.

Here we focus on magnetic atoms deposited on conducting surfaces~\cite{Hirjibehedin_Lin_Science_2007,Khajetoorians_Lounis_prl_2011,Khajetoorians_Chilian_nature_2010,Khajetoorians_Schlenk_prl_2013}, where scanning tunneling microscopes (STM) can probe the two quantities that characterize quantum or classical behavior: the quantum spin tunneling splitting $\Delta$, which can be measured by inelastic electron tunneling spectroscopy~\cite{Hirjibehedin_Lin_Science_2007,Tsukahara_Noto_prl_2009,Khajetoorians_Chilian_nature_2010}, and  their magnetization, accessible through spin polarized STM~\cite{Wiesendanger_revmod_2009}.

 It has been found that diverse magnetic adatoms can be described with the spin 
Hamiltonian:~\cite{Gatteschi_Sessoli_book_2006,Hirjibehedin_Lin_Science_2007,Khajetoorians_Chilian_nature_2010,Gruber_Drabenstedt_science_1997}
\begin{equation}
{\cal H}_{\rm S} = D \hat{S}_z^{2} + E\left(\hat{S}_x^{2}-\hat{S}_y^{2}\right). 
\label{singlespin}
\end{equation}
%
This Hamiltonian yields a type Q spectrum for integer spins $S$ with negative uniaxial anisotropy $D<0$ and  finite in-plane anisotropy $E$. 
In that case,  both the non-degenerate ground state $|\phi_G\rangle$ and the first excited state $|\phi_X\rangle$, split by $\Delta_0\propto E (E/D)^ {S-1}$, satisfy Eq. (1) with $|C_1\rangle \approx |+S\rangle$ and $|C_2\rangle \approx  |-S\rangle$ (see Fig. 1a).
This Hamiltonian correctly accounts for  the observed $dI/dV$ spectra of   Fe adatoms on Cu$_2$N/Cu(100)~\cite{Hirjibehedin_Lin_Science_2007}, Fe Phthalocyanine (FePc) molecules on CuO/Cu(110)\cite{Tsukahara_Noto_prl_2009} and Fe adatoms on InSb \cite{Khajetoorians_Chilian_nature_2010}  (with $S=2$ for Fe/Cu$_2$N and $S=1$ for the others).  In these three systems the $dI/dV$ spectra reveal finite quantum spin tunneling between $|\phi_G\rangle$ and $|\phi_X\rangle$ and a null magnetic moment can be expected.

 Spin polarized STM  magnetometry on  short chains 
of Fe atoms on Cu$_2$N/Cu(100) are not able\cite{Loth_Baumann_science_2012} to detect magnetic moment, consistent with  a type Q behavior and the observation of QST splitting on the single atom. Intriguingly, longer chains display a spontaneous atomic magnetization, in the form of antiferromagnetically aligned N\'eel states~\cite{Loth_Baumann_science_2012}.    

 This paper is devoted to understanding how 
 the conventional classical picture of a magnet with two equivalent ground states with opposite magnetization emerges for type Q magnetic adatoms.  
We discuss two independent mechanisms:
Kondo exchange,  that operates even for a single magnetic adatom, and interatomic exchange.
 The driving factor in both cases is the quenching of the QST splitting so that the dressed type Q system becomes effectively a type C exhibiting two classical degenerate ground states.

\begin{figure}[!]
  \begin{center}
    \includegraphics[width=0.8\linewidth,angle=0]{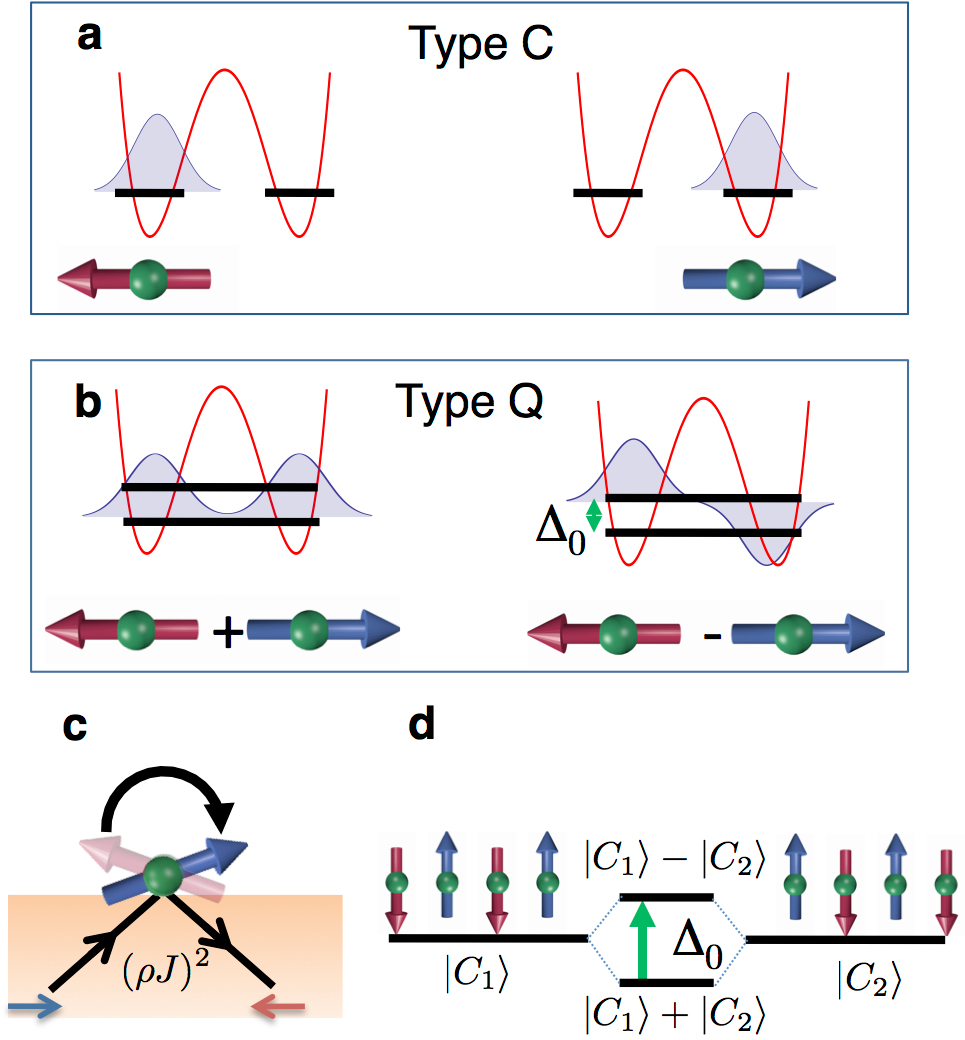} 
  \end{center}
  \caption{ {\em Two types of quantized spin systems.} (a) Scheme of a type C spin system, with an easy axis and two degenerate ground states, 
bearing each a finite magnetic moment.
(b) Scheme of a type Q spin system.  Due to QST, bonding and antibonding linear combination
of states with opposite magnetization,
are formed and are  separated in energy by $\Delta_0$, the QST splitting.
(c)  Scheme of Kondo exchange interaction between a single magnetic atom and conduction electrons  that quenches $\Delta_0$ (see Fig. 2) as $(\rho J)$, the product of the density of states of conducting electrons and the Kondo exchange,  is increased.
(d)  Representation of the two classical degenerate N\' eel states for spin chains, denoted as $|C_1\rangle$ and $|C_2\rangle$, as well as the
 type Q bonding and anti-bonding states and their corresponding QST splitting.  \label{fig1}
  }
\end{figure}

We consider first Kondo exchange in the  the weak coupling regime, where the magnetic adatom spin preserves its
identity and the Kondo singlet  has not been formed. In that limit,  perturbation theory~\cite{Delgado_Palacios_prl_2010}  predicts that  Kondo exchange produces both a broadening $\Gamma$ and a shift of the atomic spin
excitations,  in agreement with experiments~\cite{Oberg_Calvo_natnano_2013,Loth_Bergmann_natphys_2010} .
 Both quantities are proportional to the dimensionless constant  $(\rho J)^2$, the product of the density of states $\rho$ of the surface electrons and the Kondo exchange $J$.
  Here we go beyond
perturbative theory and show that  a sufficiently large $\rho J$ quenches completely the QST.  
To do so, we assume that the separation of the ground stated doublet from
the higher excited states is larger than all relevant energy scales, such as thermal energy or the QST, so  that we truncate the Hilbert space keeping only the two lowest energy
 states, $|\phi_G\rangle$ and $|\phi_X\rangle$. 
Hence, defining  
a pseudo-spin $1/2$, with Pauli matrices $\vec{\tau}$, 
the atomic spin operator represented in this space takes the form:
\beqa
\left(\hat S_x,\hat S_y,\hat S_z\right)\rightarrow \langle \phi_G|\hat S_z|\phi_X\rangle\left(0,0,\hat \tau_x\right).
\label{dos}
\eeqa
This means that the exchange coupling acts only through the $S_z$-conserving (Ising) channel, which prevents the formation of a Kondo singlet, but
creates a pseudo-spin flip in the space of $|\phi_G\rangle$ and $|\phi_X\rangle$. 
As a consequence, the Kondo Hamiltonian projected in the $(\phi_G,  \phi_X)$ subspace has the form
\begin{equation}
H_{\rm K}\equiv  \sum_{\vec{k},\sigma} \epsilon_{|\vec{k}|} c^{\dagger}_{\vec{k},\sigma} c_{\vec{k}\sigma}
+
\frac{\Delta_0}{2}\hat{\tau}_z + 
\hat\tau_x
\sum_{ \vec{k},\vec{k}'} \frac{j}{2}
\left(c^{\dagger}_{\vec{k},\uparrow} c_{\vec{k}'\uparrow}-c^{\dagger}_{\vec{k},\downarrow} c_{\vec{k}'\downarrow}\right),
\label{Kondo}
\end{equation}
where $j=J\langle \phi_G|\hat S_z|\phi_X\rangle$. This Hamiltonian is known as the Ising-Kondo model in a transverse field~\cite{Sikkema_Buyers_prb_1996}.

 For a point scatterer, conduction electrons can be described as one
dimensional fermions, which permits making use of the bosonization technique~\cite{Schotte_Schotte_prb_1969}
where  charge and spin densities are represented in terms of bosonic operators $b_k,b_k^\dag$~\cite{Mattis_Lieb_jmatphys_1965,Leggett_Chakravarty_rmphys_1987}.
This allows mapping the
original Kondo model for the type Q spins into the spin-boson (SB) Hamiltonian with
an Ohmic spectral density~\cite{Bray_Moore_prl_1982,Leggett_Chakravarty_rmphys_1987}
\beqa
H_{SB}&=&\frac{\Delta_0}{2}\hat{\tau}_z+\hbar v_F \sum_{k>0} k b_k^\dag b_k
\crcr
&&+\hbar v_F \hat\tau_x\sqrt{\pi\alpha} \sum_k\sqrt{\frac{|k|}{L}}e^{-k v_F/2\omega_c}\left(b_k^\dag +b_k\right),
 \cr &&
\label{hsb}
\eeqa
%
where $v_F$  is the Fermi velocity, $L$ is the size of the system, and 
$\hbar \omega_c$ is the bosonic energy cut-off. 
Here the first term describes the QST of the bare magnetic atom, the second  the surface electrons, and  the third term accounts for the Kondo interaction.  The constant 
$\alpha=\left(\rho J\right)^2\left|\langle \phi_G|S_z|\phi_X\rangle\right|^2$ plays a key role in the SB model (see Supplemental information for details).

\begin{figure}[t]
  \begin{center}
    \includegraphics[width=0.9\linewidth,angle=0]{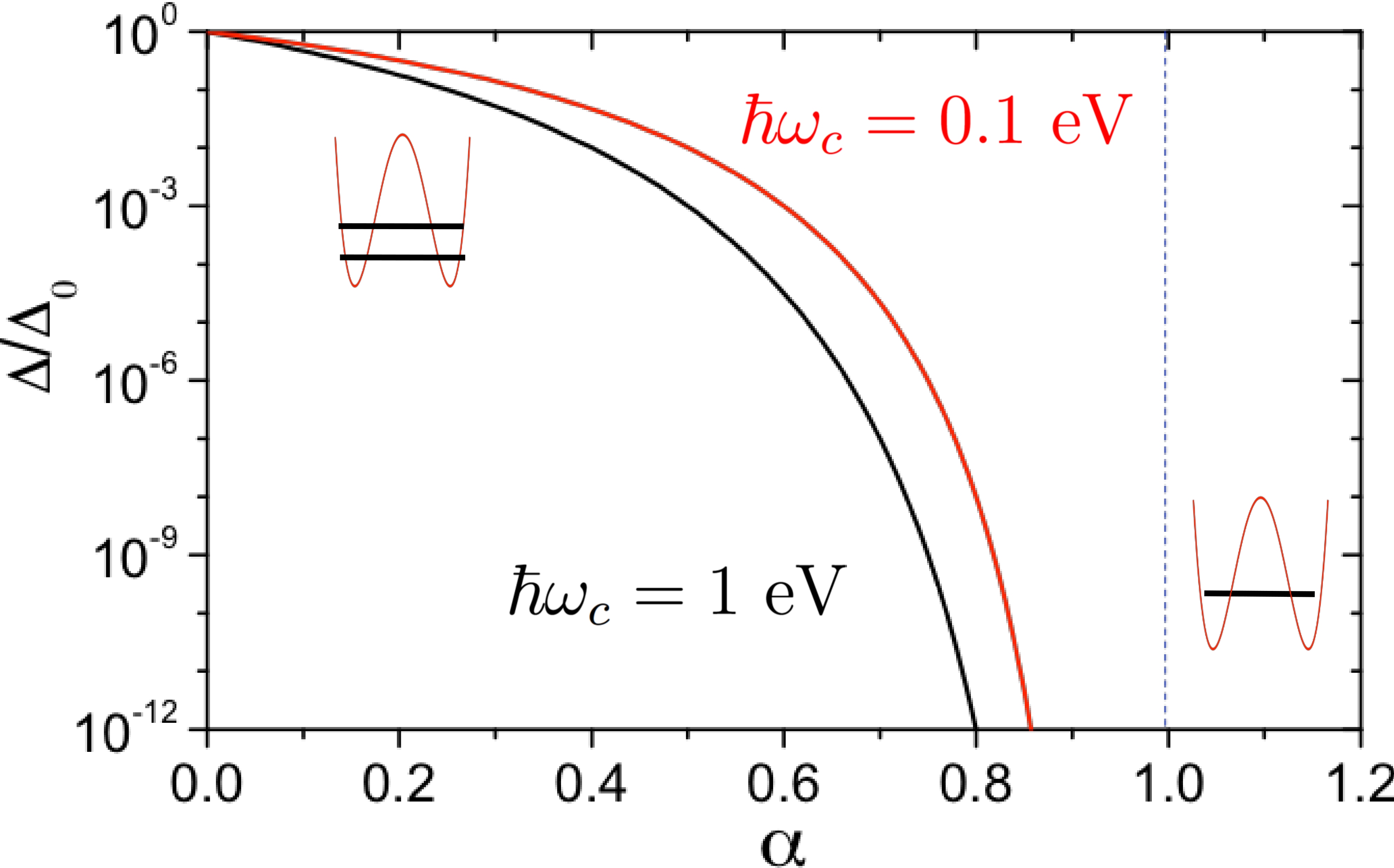} 
  \end{center}
  \caption{ {\em Renormalization of the QST splitting for a Kondo coupled spin.} Zero-temperature renormalized splitting $\Delta$ in units of the unperturbed level splitting $\Delta_0$, Eq. (5), of a type Q  single spin Kondo-coupled to an electron gas
as function of coupling strength. 
Red (Black) lines show two different energy cut-offs, $\hbar \omega_c$ ($\Delta_0=1$ meV).   $\Delta$ is very rapidly quenched (notice logarithmic scale) and 
vanishes identically for $\alpha\to 1$.   \label{fig2}
  }
\end{figure}

The SB Hamiltonian is a paradigm model  to describe a quantum to classical phase transition induced by
environmental decoherence~\cite{Leggett_Chakravarty_rmphys_1987}. The transition is driven by the competition between QST (with strength $\Delta_0$),  which favors the mixing of states with opposite $S_z$, and the
Kondo coupling (with strength $\alpha$) that favors localization of the system in one of the two states with $S_z =\pm S$. 
The SB model yields the following non-perturbative result for the renormalization of the QST
splitting $\Delta$ at zero temperature~\cite{Leggett_Chakravarty_rmphys_1987,Hur_annphys_2008}:
\beqa
\frac{\Delta}{\Delta_0}\approx\left(1-\Theta (\alpha)\right)
		       \left(\frac{\Delta_0}{\hbar \ome_c}\right)^{\frac{\alpha}{1-\alpha}},
\eeqa
%
where $\Theta$ is the step function. Hence, increasing $\alpha$ decreases $\Delta/\Delta_0$ exponentially
fast, as shown in Fig. 2, vanishing completely when $\alpha\ge 1$.
This point marks a zero-temperature quantum phase
transition beyond which quantum tunneling is suppressed.  

Importantly, IETS measurements permits inferring $\alpha$. The full width at half maximum $W_{2,1}$ of the peak (or dip) in the experimental $d^2I/dV^2$, corresponding to the $|\phi_1\rangle \to |\phi_2\rangle$ transition, is related to the 
intrinsic relaxation rate $\Gamma_{2,1}/\hbar$  via
\cite{Lauhon_Ho_revsciin_2001}
\beqa
\label{width}
\Gamma_{2,1}=\sqrt{ e^2W_{2,1}^2-\left(1.7eV_{AC}\right)^2-\left(5.4k_BT\right)^2},
\eeqa
where  $V_{AC}$ is the modulation voltage in the applied bias. In addition, and within perturbation theory, the exchange-induced energy broadening $\Gamma_{2,1}$ is given by~\cite{Delgado_Palacios_prl_2010} 
\beqa
\Gamma_{2,1}=\alpha \frac{\pi}{2}\Delta_{21}\left[1+n_B(\Delta_{21})\right]
\frac{\sum_a\left|\langle\phi_1|\hat S_a|\phi_2\rangle\right|^2}{\left|\langle \phi_G|S_z|\phi_X\rangle\right|^2},
\label{gamma}
\eeqa
%
where $\Delta_{21}$ is the energy of the transition and $n_B$ is the thermal Bose factor.
Thus,  from the experimental results 
we find that, while  Fe on Cu$_2$N/Cu(100) has $\alpha<0.1$ and a finite $\Delta$,  Fe  on Cu(111) has $\alpha\approx 2$,  
leading to $\Delta = 0$ even if symmetry-breaking effects or higher-order anisotropy contributions  remove the degeneracy caused by the $C_{3v}$ symmetry~\cite{Miyamachi_Schuh_nature_2013,Khajetoorians_Schlenk_prl_2013} of the surface.
%
%

\begin{figure}[t]
  \begin{center}
    \includegraphics[width=0.99\linewidth,angle=0]{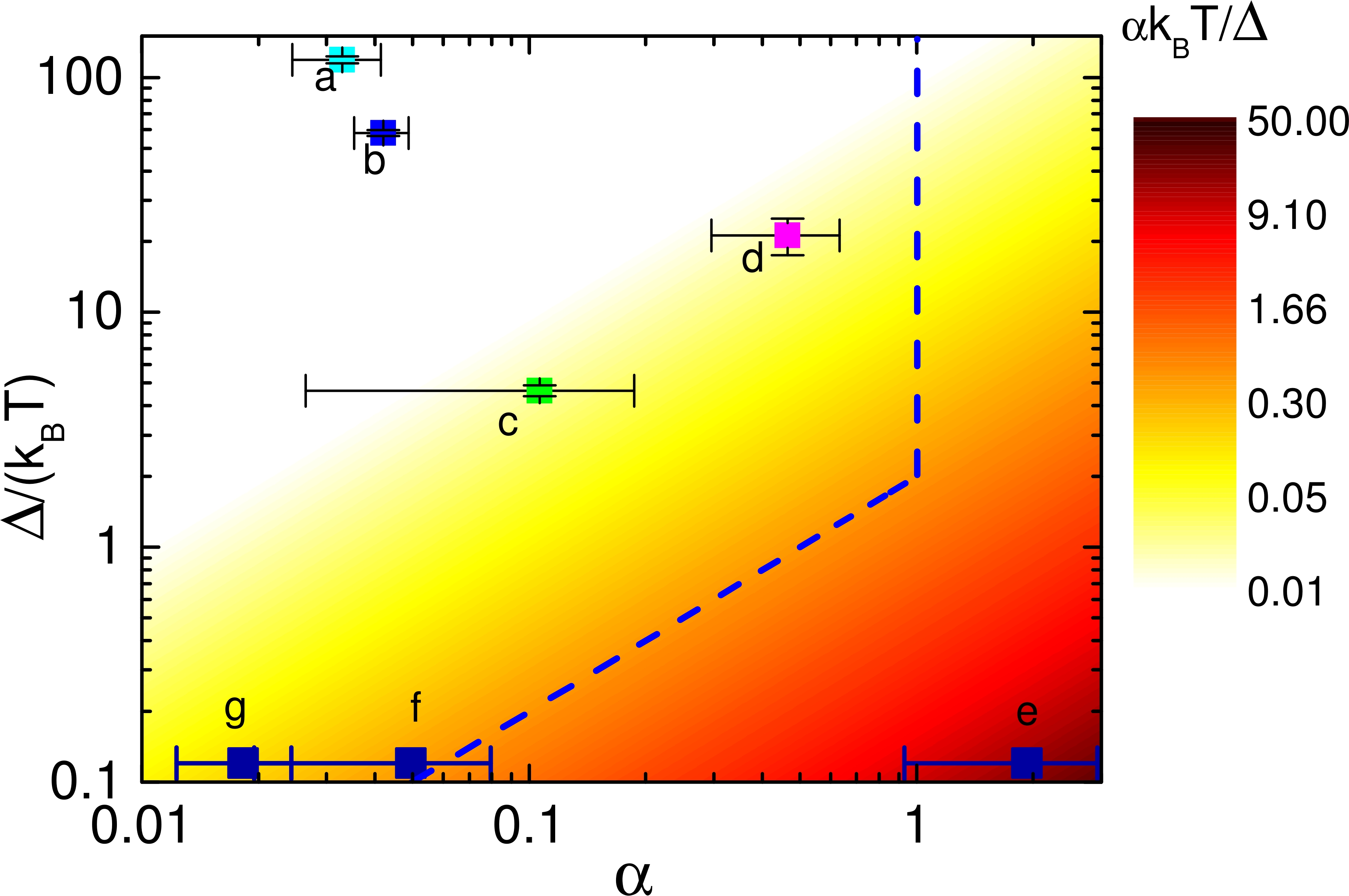} 
  \end{center}
  \caption{{\em Finite-temperature phase diagram.}
The phase space spanned by the 
renormalized QST splitting $\Delta$ in units of $k_B T$
and the substrate coupling constant, $\alpha$, is classified into {\em Quantum} (where $\alpha k_B T/\Delta\ll 1$),
 and {\em Classical} ($\alpha k_B T/\Delta\gg 1$) 
regimes. The blue dashed
line indicates the limiting condition $\alpha={\rm Min}\left[\Delta/(2k_B T),1\right]$.
Experimentally reported type Q spin systems: (a),(b) Fe-Phthalocyanine molecules on
CuO/Cu(110)~\cite{Tsukahara_Noto_prl_2009},
 (c) Fe atoms on
Cu$_2$N/Cu(100)~\cite{Hirjibehedin_Lin_Science_2007},
and (d) Fe dopants on
InSb(110)~\cite{Khajetoorians_Lounis_prl_2011}.
Measured type C systems, where $\Delta$ could not be determined experimentally,  
  are shown over the horizontal axis:
(e) Fe atoms on Cu(111)
~\cite{Khajetoorians_Chilian_nature_2010},
and (g) and (f) Fe atoms on Pt(111)~\cite{Khajetoorians_Schlenk_prl_2013}.
  \label{fig3}
  }
\end{figure}

In the SB model at finite temperature, spin Rabi oscillations, and therefore the existence of non-degenerate superposition ground states, are suppressed when $\Delta/(k_BT) \lesssim  2\alpha$ \cite{Leggett_Chakravarty_rmphys_1987}.  This enables to place the experimental observations in a phase diagram $(\alpha, \Delta/k_BT)$, shown in Fig. 3,  separating quantum and classical regions.

We now address the emergence of 
classical behaviour in chains of spin-coupled atoms. Sufficiently long chains of Fe atoms present classical behaviour (type C) with two degenerate N\'eel states, whereas isolated Fe atoms show QST~\cite{Loth_Baumann_science_2012}.
 Here Fe-Fe exchange causes a renormalization of the QST splitting  even in the absence of Kondo-coupling to the substrate conduction electrons. We use the following Hamiltonian for $N$ spins:
 \begin{equation}
 {\cal H}=\sum_{n=1}^N {\cal H}_{\rm S}(n) + J_{\rm H} \sum_{n=1}^{N-1} \vec{S}(n)\cdot\vec{S}(n+1),
 \label{Heis}
 \end{equation}
%
where the first term describes the single ion Hamiltonian of Eq. (2) for each Fe, and the second
their antiferromagnetic exchange ($J_H>0$). When acting independently, both terms yield a unique
ground state without spontaneous magnetization. However, their combination gives non-trivial
results. This can be first seen using the same truncation scheme of the single atom case, keeping
only 2 levels per site. Hamiltonian (7) then maps into the quantum Ising model with a transverse field
(QIMTF) (more details in Supplementary information):
\begin{equation}
 {\cal H}\equiv \sum_{n=1}^{N}  \frac{\Delta_0}{2} \hat \tau_z(n) + 
j_H\sum_{n=1}^{N-1} \hat\tau_x(n)\hat \tau_x(n+1)
\label{Ising}
\end{equation}
%
where $j_H=J_H|\langle \phi_G|\hat S_z|\phi_X\rangle|^2$.  This model can be solved exactly and presents a quantum phase
transition in the thermodynamic limit ($N\rightarrow \infty$), separating a type C from a type Q phase.
In terms of the dimensionless parameter $g\equiv 2j_H/\Delta_0$, the transition occurs at $g_c=1$~\cite{Pfeuty_annalp_1970}.
 For $g<1$ the spin chain is in a quantum paramagnetic phase with
a unique ground state and $\langle \hat \tau_x(n)\rangle=\langle \hat S_z(n)\rangle =0$. For $g\ge 1$, it is in a magnetically ordered phase,
with 2 equivalent ground states with staggered magnetization, $\langle \hat \tau_x(n)\rangle \propto \langle \hat S_z(n)\rangle \propto(-1)^n$. In this 
thermodynamic limit, the QST is renormalized by the  interactions according
to~\cite{Pfeuty_annalp_1970}
\beqa
\label{rqim}
\frac{\Delta}{\Delta_0}=\left|1-g\right|\Theta(1-g).
\eeqa
%
This result shows that, as in the case of Kondo exchange in Eq. (5), interatomic exchange
also renormalizes QST, and when sufficiently strong, suppresses it completely. However, direct application  of 
Eqs. (8) and (9) 
is not possible for systems where exchange and anisotropy are of the same order,
 $J_H\sim |D|$, preventing the use of the mapping to an Ising model.  
Instead, we compute the eigenstates of Hamiltonian (7)
numerically, and compare with those of finite size chains of the QIMTF (Fig. 4). In both systems we find the same phenomenology: both the ground and first excited states, $|\phi_G\rangle$ and $|\phi_X\rangle$ , satisfy Eq. (1) with $|C_1\rangle$ and $|C_2\rangle$ being classical N\'eel-like states, and the next excited states lie much higher in energy.   Importantly,  it is still true that the QST splitting is renormalized by the interatomic exchange (inset of Fig. 4a).  Crucially, for the observed value $J_H=0.7$meV~\cite{Loth_Baumann_science_2012,Bryant_Spinelli_prl_2013},  the quenching of $\Delta$ increases exponentially with the size of the chain (Fig. 4b).

\begin{figure}[!]
  \begin{center}
    \includegraphics[width=0.9\linewidth,angle=0]{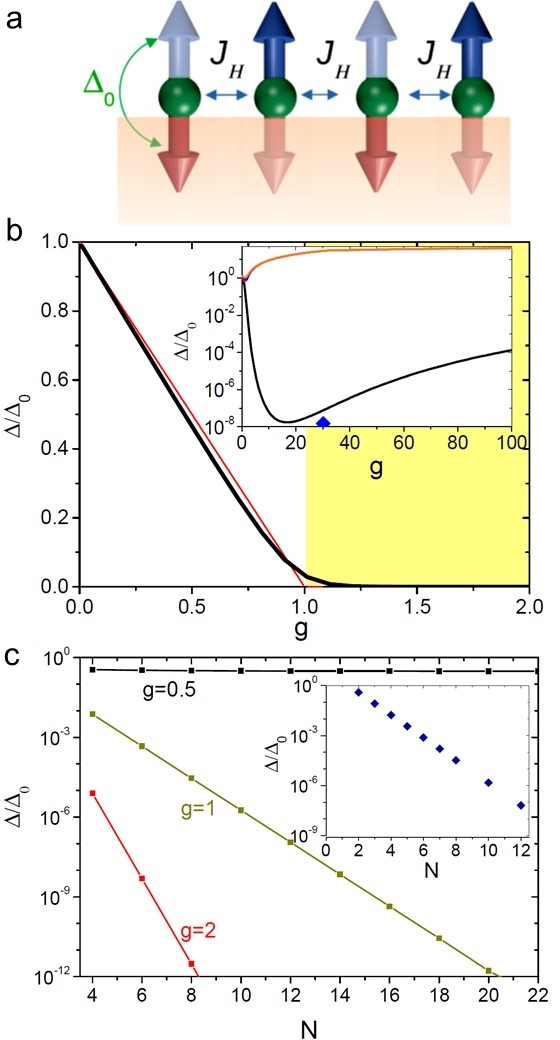} 
  \end{center}
  \caption{  {\em Quenching of quantum spin tunneling in spin chains.} (a) Schematic of a superposition state of two N\' eel states in a finite
 antiferromagnetic spin chain.
(b) QST splitting of a $S=1/2$ Ising chain in a transverse field, Eq. (8), versus the dimensionless exchange coupling $g=2j_H/\Delta_0$ for a finite $N=20$ chain (black line) and the infinite chain (red). The infinite chain has a quantum phase transition to the classical (shaded) region at $g=1$. The inset shows the QST splitting of the $S=2$ Heisenberg spin chain together with higher energy excitations (orange lines) versus exchange coupling for Fe chains with $D=-1.5$ meV and $E=0.3$ meV~\cite{Loth_Baumann_science_2012,Bryant_Spinelli_prl_2013} (the diamond marks the experimental condition where $g\approx 27$). (c) Chain size dependence of $\Delta$ in the QIMTF for Ising  spin coupling, for $g=0.5<g_c$ (weak size dependence),  and $g=1,\;2$ (exponential dependence that leads to a type Q ground state for large $N$). Inset:  size dependence of $\Delta$ for Hamiltonian (\ref{Heis}) with the experimental parameters~\cite{Loth_Baumann_science_2012,Bryant_Spinelli_prl_2013},
showing an exponential dependence analogous to the Ising case with $g>1$ (type C system).   \label{fig4}
  }
\end{figure}

In rigor, interatomic exchange in finite chains renormalizes the QST to a tiny but finite value (see Fig.4). Therefore, the observed~\cite{Loth_Baumann_science_2012} emergence of classical behavior is  
probably assisted as well by the Kondo coupling. 
Using results from second order perturbation theory, one finds that the Kondo-induced decoherence rate of a chain of $N$ spins is  
\beqa
T_2^{-1}(N)= \frac{N\pi}{2}\alpha S^2\frac{k_BT}{\hbar}.
\label{T2}
\eeqa
For instance, the $N=8$ Fe chain of Ref.~\cite{Loth_Baumann_science_2012}
leads to
$\Delta /(\hbar T_2^{-1}) \lesssim  10^{-6}$ at $T=0.5$ K,  indicating that the Fe chain will be in the decohered type C state.
Thus, the combination of interatomic exchange, that reduces almost down to zero the QST of the monomer, and the enhanced spin decoherence of the chain due to Kondo exchange with the substrate, lead to the emergence of classical behaviour of the finite size spin chains.

Our results provide a general scenario for the emergence of classical magnetism in quantum spins systems that have a unique ground state superposition of classical states with opposite magnetizations. A sufficiently strong coupling to either the itinerant electrons or to other localized spins,  leads to a phase with a doubly degenerate ground state where classical behaviour appears. Using experimentally verified coupling strengths we find that this transition can occur for small ensembles of interacting atoms ($N<10$) or even individual atoms.
Hence, the classical phase in nanomagnets appears as a quantum phase transition to a quantum decohered phase.



{\bf Acknowledgements}
 We acknowledge A. J. Heinrich, R. Aguado, M. A. Cazalilla and  A. Khajetoorians for fruitful discussions. This work has been financially supported by 
  Generalitat Valenciana, grant Prometeo 2012-11.

\end{document}